\documentclass[a4paper,11pt]{article}
\usepackage{pos}
\usepackage{hyperref}

\title{Particle spectra from dark matter annihilation: physics modeling and QCD uncertainties}
\ShortTitle{Particle spectra from dark matter annihilation}

\author[a]{Simone Amoroso}
\author[b,c]{Sascha Caron}
\author*[d]{Adil Jueid}
\author[e]{Roberto Ruiz de Austri}
\author[f]{Peter Skands}

\affiliation[a]{Deutsches Elektronen-Synchrotron (DESY), Notkestrasse 85, D-22607 Hamburg, Germany}
\affiliation[b]{Institute for Mathematics, Astrophysics and Particle Physics, Faculty of Science, Mailbox 79,
Radboud University Nijmegen, P.O. Box 9010, NL-6500 GL Nijmegen, The Netherlands}
\affiliation[c]{Nikhef, Science Park, Amsterdam, The Netherlands}
\affiliation[d]{School of Physics, Konkuk University, 05029, Seoul, Republic of Korea}
\affiliation[e]{Instituto de F\'isica Corpuscular, IFIC-UV/CSIC, Valencia, Spain}
\affiliation[f]{School of Physics and Astronomy, Monash University, VIC-3800, Australia}

\emailAdd{simone.amoroso@desy.de}
\emailAdd{scaron@cern.ch}
\emailAdd{adil.hep@gmail.com}
\emailAdd{rruiz@ific.uv.es}
\emailAdd{peter.skands@monash.edu}

\abstract{In this talk, we discuss the physics modelling of particle spectra arising from dark matter (DM) annihilation or decay. In the context of the indirect searches of DM, the final state products will, in general, undergo a set of complicated processes such as resonance decays, QED/QCD radiation, hadronisation and hadron decays. This set of processes lead to stable particles (photons, positrons, anti-protons, and neutrinos among others) which travel for very long distances before reaching the detectors. The modelling of their spectra contains some uncertainties which are often
neglected in the relevant analyses. We discuss the sources of these uncertainties and estimate their impact on photon energy spectra for benchmark DM scenarios with $m_\chi \in [10, 1000]\,$GeV. Instructions for how to retrieve complete tables from Zenodo are also provided. 
}

\FullConference{%
  Tools for High Energy Physics and Cosmology - TOOLS2020\\
  2-6 November, 2020\\
  Institut de Physique des 2 Infinis (IP2I), Lyon, France}

\begin{document}
\maketitle

\section{Introduction}
\label{sec:introduction}

Various gravitational, astrophysical and cosmological observations strongly imply the existence of Dark Matter (DM) in the universe. In particular, observations of the cosmological scale structure favour the so-called cold DM (CDM) scenario where the DM was not relativistic in the era of structure formation. In particle physics framework, the CDM scenario can be easily accounted for by extending the Standard Model (SM) with weakly interacting massive particles (WIMPs) -- for a review see e.g. \cite{Bertone:2004pz} --. An interesting feature of the CDM scenario is that WIMPs with mass about $\mathcal{O}(100)~$GeV interacting primarily through weak interactions gives relic abundance of DM in agreement with the observation made by the Planck satellite, \emph{i.e.} $\Omega_{\mathrm{DM}} h^2 = 0.1188 \pm 0.0010$ \cite{Ade:2015xua}. \\

Indirect detection experiments such as the Fermi Large Area Telescope (LAT) \cite{Atwood:2009ez}, AMS \cite{Aguilar:2013qda} or IceCube \cite{Aartsen:2012kia} provide one the possible ways to detect WIMPs. Theoretically, WIMPs undergo either annihilation \cite{Griest:1990kh, Bergstrom:2000pn}, co-annihilations \cite{Baker:2015qna}, or decays \cite{Cata:2016dsg, Azri:2020bzl} into a set of SM stable final state particles such as high energy photons, positrons, neutrinos, or anti-protons. Recently, an excess on the gamma-ray spectra was detected by the Fermi-LAT \cite{TheFermi-LAT:2015kwa}, called the Galactic Center Excess (GCE) which apparently seem to be consistent with predictions from DM annihilation (see e.g. \cite{Goodenough:2009gk}). On the other hand, several attempts were made to address the GCE within particle physics models, in particular within supersymmmetric models \cite{Caron:2015wda, Bertone:2015tza, Butter:2016tjc, Achterberg:2017emt}. An important finding is that the quality of the fits depend crucially on the theoretical precision on the determination of the gamma-ray spectra \cite{Caron:2015wda}.  \\

Particle production from DM annihilation/decay processes is dominated by QCD jet fragmentation.\footnote{This is true for DM masses above a few GeV producing hadronic final states either directly through e.g. $\chi\chi \to q\bar{q}$ or indirectly via the decays of the intermediate heavy resonances such as the $W/Z/H$ bosons or the top quark.} Final state stable particles such as photons or positrons are then produced as a result of a complicated set of processes which includes QED and QCD radiations, hadronisation, and hadron decays. Unlike parton-level scattering amplitudes at e.g. the lowest order of perturbation theory, the problem of hadronisation cannot be solved from first-principles. Jet universality tells us that hadronisation is a universal process that can be factorised off the short-distance processes e.g. DM annihilation. Phenomenological models such that Fragmentation Functions \cite{Metz:2016swz} or explicit dynamical models such as the string~\cite{Artru:1974hr,Andersson:1983ia} or cluster~\cite{Webber:1983if,Winter:2003tt} models which are embedded in Monte Carlo (MC) event generators~\cite{Buckley:2011ms} are the up-to-date solutions to the hadronisation problem. The essential point is that the fragmentation models' parameters are to a very good approximation independent of the short-distance processes; therefore, they can be determined from fits to existing data such as $e^+ e^- \to \mathrm{hadrons}$ and used to make predictions for e.g. DM annihilation. \\

The question of the intrinsic QCD uncertainties on the predicted particle spectra in DM annihilation is often neglected in the literature besides some comparisons between the predictions of different multi-purposes event generators such as \textsc{Herwig} and \textsc{Pythia}. For instance, a comprehensive analysis has shown that different MC event generators may have excellent agreement in both the peak as well as the bulk of the spectra while their agreement is not very good in the tails \cite{Cembranos:2013cfa}. Another study was done by the authors of the PPPC4DMID \cite{Cirelli:2010xx} where they highlighted the differences between \textsc{Herwig} and \textsc{Pythia} event generators. The excellent level of agreement in the most-populated regions of particle spectra 
may be interpreted as due to the fact that the different MC generators tend to be tuned to roughly the same set of data mostly coming from LEP measurements at the $Z$-boson pole \cite{Buckley:2009bj,Buckley:2010ar,Skands:2010ak,Platzer:2011bc,Karneyeu:2013aha,Skands:2014pea,Fischer:2014bja,Fischer:2016vfv,Reichelt:2017hts,Kile:2017ryy}. Therefore, the envelope spanned by the predictions of the different MC models cannot represent the true estimate of the uncertainty on the predicted spectra. \\

In this talk, we discuss a \emph{first} study of the QCD uncertainties on particle spectra from DM annihilation
within the same MC model\footnote{In this work, we focused on the uncertainties within the \textsc{Pythia8} event generator and the results we shown are based on \cite{Amoroso:2018qga}.}. We take the default Monash 2013 tune~\cite{Skands:2014pea} of the \textsc{Pythia} version 8.2.35 event generator~\cite{Sjostrand:2014zea} as our baseline. We use a selection of experimental measurements constraining from $e^+e^-$ colliders preserved in the \textsc{Rivet}~\cite{Buckley:2010ar} analysis package combined with the \textsc{Professor}~\cite{Buckley:2009bj} parameter optimisation tool. Then, we define a small set of the systematic parameter variations which we argue that explores the uncertainty envelope for the estimate of the QCD fragmentation uncertainties on DM annihilation. 

\begin{figure}[!t]
\centering
\begin{tabular}{c|cc}
QED bremsstrahlung & \multicolumn{2}{c}{QCD fragmentation and hadron decays}\\
\includegraphics[scale=0.65]{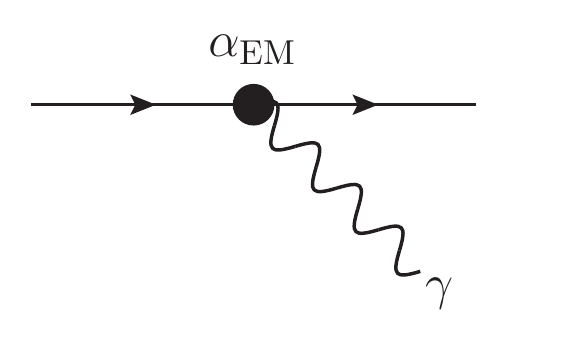} &
\includegraphics[scale=0.65]
{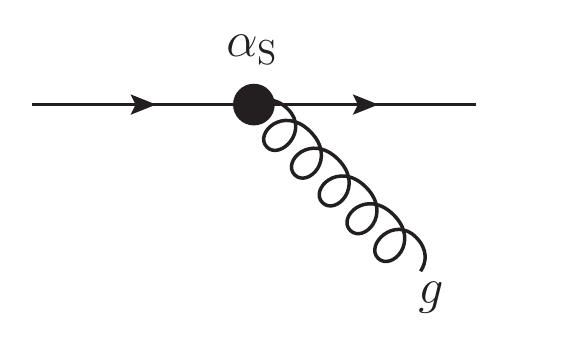} &
\raisebox{-2.5mm}{\includegraphics[scale=0.65]{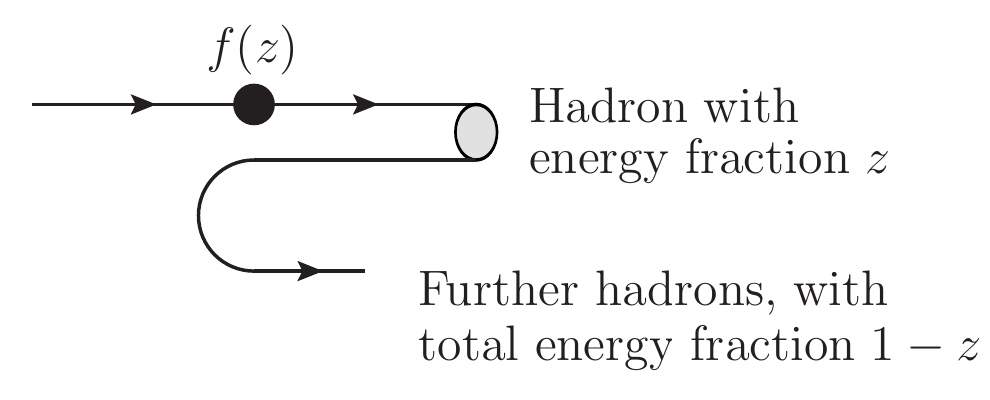}}\\
Dominates at high $x_\gamma$ & \multicolumn{2}{c}{Photons from $\pi^0\to\gamma\gamma$ dominate bulk (and peak) of spectra}
\end{tabular}
\caption{Illustration of the main parameters that affect the photon spectra ($x_\gamma = E_\gamma/m_\chi$) from DM annihilation into jets. Here we show the electromagnetic coupling $\alpha_\mathrm{EM}$ (left), the strong coupling $\alpha_S$ (middle), and the nonperturbative fragmentation function $f(z)$ (right). \label{fig:MCparams}}
\end{figure}

\section{Physics Modeling and Measurements}
\label{sec:physics-modeling}
\subsection{Physics modeling}
In this section, we discuss briefly the physics modeling in a generic DM annihilation, and the origin of photons (a more detailed discussion can be found in \cite{Amoroso:2018qga}). To simplify the discussion, we consider a generic DM annihilation process:
\begin{eqnarray*}
 \chi \chi \to X_1 \cdots X_n \to \displaystyle\prod_{i=1}^{m} Y_{1i} \cdots Y_{ni}
\end{eqnarray*}
where we factorised the whole process into a production part $\chi \chi \to X_1 \cdots X_n$, and the decay part ($X_i \to Y_{i1} \cdots Y_{in}$ with $Y_{ij}$ is any stable object such as photon, neutrino or proton) assuming the narrow-width approximation. We note three important processes which may occur after DM annihilation and responsible for gamma-rays:

\begin{itemize}
    \item \emph{QED bremsstrahlung:} This process occurs if $X$ (or the decay products $Y$) contains photons or electrically charged particles. Additional photons are produced via $X_i^\pm \to X_i^\pm\gamma$ branchings with probabilities that are enhanced for both soft and collinear photons. On the other hand, collinear photons dominate the spectra at the region $x_\gamma \to E_\gamma/m_\chi \to 1$ with the only requirement that the angle between the emitted photon and the parent particle is very small. QED processes may lead to the production of charged fermion-antifermion pairs in photon splittings (which are generally subleading) and the corresponding probabilities are enhanced at very low values of $Q^2 = (p_f + p_{\bar{f}})^2$. The rates of QED processes are governed by the effective electromagnetic fine-structure constant, $\alpha_{\mathrm{EM}}$ (illustrated in Fig.~\ref{fig:MCparams}a).

    \item \emph{QCD showers:} If $X$ (or decay products $Y$) contains coloured particles, then these states will undergo QCD showers. The modeling of the QCD showers is similar to the QED one. Here, we can have enhancement of soft and collinear emissions -- in $q\to qg$ and $g\to gg$ -- and of $g\to q\bar{q}$ at low virtualities. The main parameter governing the QCD showering is the effective value of the strong coupling constant, $\alpha_S$ (see Fig. \ref{fig:MCparams}b) evaluated at a scale proprtional to the shower evolution variable ($p_\perp$ in \textsc{Pythia8}). Further sets of universal corrections in the soft limit implies that the strong coupling should be defined in the CMW~\cite{Catani:1990rr} rather than the conventional $\overline{\mathrm{MS}}$ scheme. Furthermore, good agreement between \textsc{Pythia8} and experimental measurements of $e^+ e^- \to 3~\mathrm{jets}$ \cite{Skands:2010ak,Skands:2014pea} increases the value of $\alpha_S(M_Z)$ by about $10\%$. Perturbative uncertainty estimates can be performed by variation of the evolution of the renormalisation by a factor of $2$ in each direction with respect to the nominal scale choice. The framework of the automated scale variations was recently implemented in \textsc{Pythia}~\cite{Mrenna:2016sih} implies a compensation of second-order terms which reestablishes the agreement with the CMW scheme. Variations of the no-universal (no-singular) components of the Dokshitzer-Gribov-Lipatov-Altarelli-Parisi (DGLAP) splitting kernels can be performed in this framework as it is detailed in \cite{Mrenna:2016sih}.
    
    \item \emph{Hadronisation and hadron decays:} Any produced coloured particles must be confined inside colourless hadrons. This process --- \emph{hadronisation} --- takes place at a distance scale of order the proton size $\sim 10^{-15}$m and in \textsc{Pythia} is modelled by the Lund string model; see~\cite{Andersson:1983ia} for details. The most majority of photons are produced from the decays of neutral pions where the number and energy of these photons are strongly correlated with the predicted pion spectra. The description of this process is embedded in the \emph{fragmentation function}, $f(z)$, which gives the probability for a hadron to take a fraction $z \in [0,1]$ of the remaining energy at each step of the (iterative) string fragmentation process (see Fig.~\ref{fig:MCparams}c). The fragmentation function $f(z)$ cannot calculated from first principles but its form can be constrained by requirements such as causality. The general form can be written as \begin{equation}
    f(z,m_{\perp h}) = N \frac{(1-z)^a}{z}\exp\left(\frac{-b m_{\perp h}^2}{z}\right)~,
    \label{eq:fz}
    \end{equation}
    where $N$ is a normalisation constant that guarantees the distribution to be normalised to unit integral, and $m_{\perp h}= \sqrt{m_h^2 + p_{\perp h}^2}$ is called the ``transverse mass'', with $m_h$ the mass of the produced hadron and $p_{\perp h}$ its momentum transverse to the string direction, $a$ and $b$ are tunable parameters which will be denoted respectively by \texttt{StringZ:aLund} and \texttt{StringZ:bLund}. We note that the $a$ and $b$ parameters are extremely highly correlated. This makes it meaningless to assign independent $\pm$ uncertainties on them. To address this question, we implement an alternative parametrisation of $f(z)$ where $b$ is replaced by a $<z>$ which represents the average $z$ fraction taken by $\rho$ mesons.  
    \begin{equation}
    \left<z_\rho\right> = \int_0^1 \mathrm{d}z \ z f(z,\left<m_{\perp\rho}\right>)~,
    \label{eq:zrho}
    \end{equation}
    which we solve (numerically) for $b$ at initialisation when the option \texttt{StringZ:deriveBLund = on} is selected in \textsc{Pythia} 8.235, using the following parameters:
    \begin{eqnarray}
    \left<m_{\perp\rho}\right>^2 & = & 
    m^2_\rho + 2( \mbox{\texttt{StringPT:sigma}})^2~,
    \\
    \left<z_\rho\right> & = &\mbox{\texttt{StringZ:avgZLund}}~.
    \end{eqnarray}
    \end{itemize}


\subsection{Photon origins and available measurements}

Here, we discuss briefly the origin of photons from DM annihilation (a very detailed discussion can be found in \cite{Amoroso:2018qga}). Most of the photons are coming from pion decays; about $88$-$95\%$ depending on the annihilation channel and on the DM mass. The contribution from $\eta$ decays is somewhat subleading which is about $4\%$. Finally, very sub-leading contributions are coming from bremsstrahlung photons and dominates in the high tail of the spectrum. Since the majority of photons ($\simeq 95\%$) are coming from pion decays, the QCD uncertainties on photon spectra is strongly correlated to those on the pion spectra. We can distinguish between primary pions directly produced from QCD fragmentation and secondary pions coming from the decays of heavier hadrons and $\tau$ leptons. In all the final states \cite{Amoroso:2018qga}, the number of secondary pions is larger than primary ones. The secondary pions account for about $70\%$-$85\%$ of the total pions. We note that the secondary pions mainly come from five sources: $\rho^\pm, \eta, \omega, D^{0,\pm}$ and $K_{S,L}$. \\

After discussing the origin of photons in DM annihilation, we conclude that in addition to the direct measurements of the photon spectrum, other measurements can be used to constrain the spectrum: \emph{(i)} the spectrum of neutral pions ($\pi^0$) since they are the most dominant source of photons in QCD jets, \emph{(ii)} the spectra of charged pions due to the fact that their number is related to $\pi^0$ by isospin symmetry and \emph{(iii)} $\eta$ spectrm as they are the second-most important source of photons in QCD jets. Finally, it is important to ensure that these tunings do not produce large corrections to infrared and collinear safe observables such as e.g. the $C$-parameter. The tunings will include the full range of these observables including the back-to-back regions which are extremely sensitive to non-perturbative QCD effects. These measurements provides important constraints on the \texttt{StringPT:sigma} parameter in particular. \\

\begin{figure}[!t]
\includegraphics[width=0.495\linewidth]{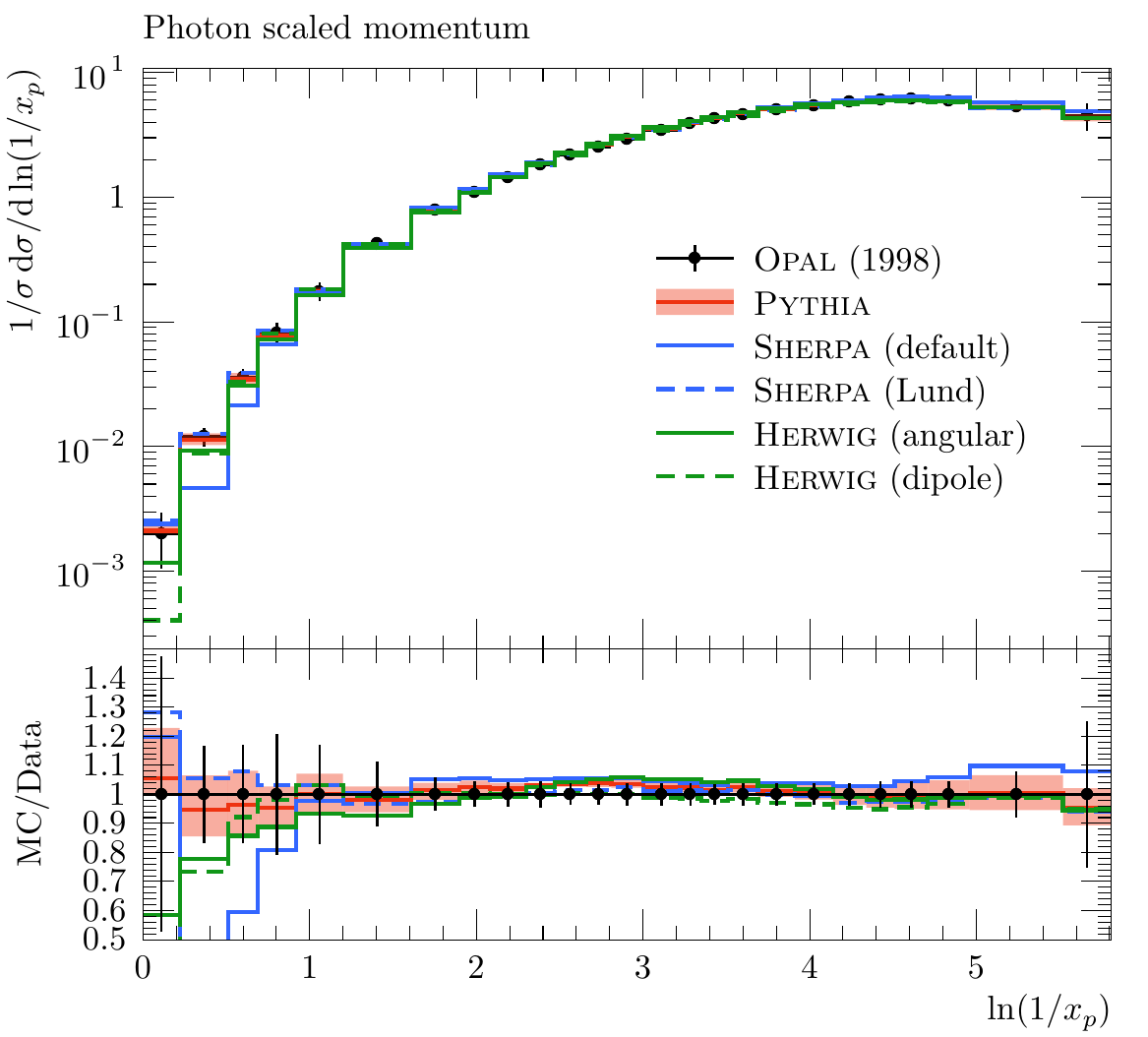}
\hfill
\includegraphics[width=0.495\linewidth]{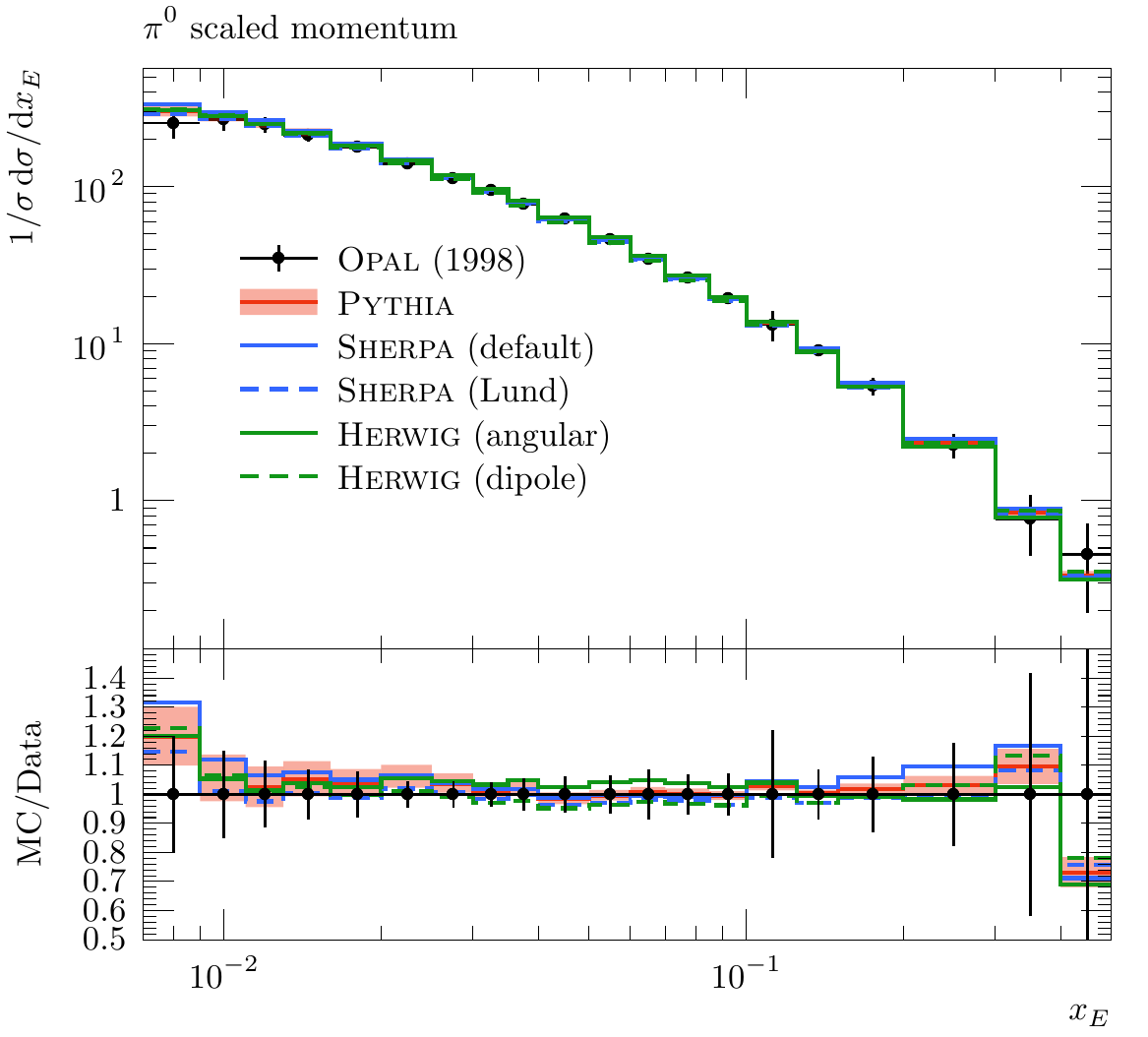}
\caption{Comparison between MC event generators and LEP and SLD measurements for the photon spectrum (left pane), and 
the $\pi^0$ spectrum (right pane). }
\label{comparison-1}
\end{figure}

In Fig. \ref{comparison-1}, we compare several different multi-purpose MC event generators to measurements of two the photon and $\pi^0$ scaled momenta. We consider three event generators in these comparisons; \textsc{Herwig} 7.1.3 \cite{Bellm:2015jjp} using both the angular-ordered \cite{Gieseke:2003rz} and dipole \cite{Platzer:2009jq, Platzer:2011bc} shower algorithms and a cluster based hadronisation model~\cite{Webber:1983if}, \textsc{Pythia} 8.2.35 with the default
model of hadronisation \cite{Sjostrand:2014zea} and \textsc{Sherpa} 2.2.5 \cite{Gleisberg:2008ta} with the CSS parton shower \cite{Schumann:2007mg} using both the Ahadic~\cite{Winter:2003tt} (based on the cluster model) and the \textsc{Pythia} 6.4 Lund hadronisation~\cite{Sjostrand:2006za} models. The curve corresponding to \textsc{Pythia} is shown with an  uncertainty band (red) obtained using the results of our new tune, based on the recent \textsc{Monash} tune but refitting the three main hadronisation parameters (see below). We can see from Fig. \ref{comparison-1} that the multi-purpose event generators agree pretty well except in a few regions such as e.g. in the tails towards hard high-energy photons. 

\section{Tuning}
\label{sec:tune}
\subsection{Setup}
We used \textsc{Pythia8} version 8.235 throughout this study with the most recent Monash~\cite{Skands:2014pea} tune is used as baseline for the  parameter optimisation. We use \textsc{Professor} v2.2~\cite{Buckley:2009bj} to perform the tuning and \textsc{Rivet} v2.5.4~\cite{Buckley:2010ar} for the implementation of the measurements. In \textsc{Professor}, a method permits to make simultaneous optimisation of several parameters by using analytical approximations of the dependence of the MC response on the model parameters (this idea was introduced first in Ref.~\cite{Abreu:1996na}). In order to minimises the differences between the interpolated functions and the true MC response, we use a fourth-order polynomial. The values of the model parameters at the minimum are then obtained with a standard $\chi^2$ minimisation of the analytic approximation to the corresponding data using \textsc{Minuit}~\cite{James:1975dr}. In this work, we tuned the $a$ and $b$ parameters of the Lund fragmentation function ($a$ and $<z_\rho>$ in the new parametrization) and the $\sigma$ parameter which governs the transverse components (see e.g. \cite{Skands:2012ts}). The default values of the parameters and their allowed range in \textsc{Pythia8} are shown in Table \ref{tab:ranges}. \\

To protect against over-fitting effects and as a baseline sanity limit for the achievable accuracy, we introduce an additional $5\%$ uncertainty 
on each bin and for each observable. This also substantially reduces
the value of the goodness-of-fit measure so that the resulting
$\chi^2/\textrm{ndf}$ is consistent with unity (see Table \ref{tab:T2tune}). The $\chi^2/N_\textrm{DoF}$ is defined by:
\begin{equation}
 \frac{\chi^2}{N_\textrm{DoF}} = \frac{1}{\sum_{\mathcal{O}} \omega_\mathcal{O} |b \in \mathcal{O}|}\frac{\sum_{\mathcal{O}}
 \omega_\mathcal{O} \sum_{b\in \mathcal{O}} (f_{(b)}(\textbf{p}) - \mathcal{R}_b)^2}
 {(\Delta_b^2+ (0.05 f_{(b)}(\textbf{p}))^2)}~.
\label{Gof-Ndf}
 \end{equation}
 
Here $\omega_\mathcal{O}$ represents the weight per observable and per bin, $f_{(b)}(\textbf{p})$ is the interpolated function per bin $b$, $\mathcal{R}_b$ is the experimental value of the observable
$\mathcal{O}$ and $\Delta_b$ is the experimental error per bin, with $f_{(b)}$ is the $4$th order interpolated polynomial used to model the MC response. We use various experimental measurements from \textsc{Lep} and \textsc{Slc} at the $Z$-boson peak produced by \textsc{Aleph},
\textsc{Delphi}, \textsc{L3}, 
\textsc{Opal}  and \textsc{Sld}.

\begin{table}[t!]
  \begin{center}
    \begin{tabular}{llcl}
      \hline
      parameter & \textsc{Pythia8} setting & Variation range & \textsc{Monash}\\
      \hline
      $\sigma_{\perp}$~[GeV] & \verb|StringPT:Sigma|   & 0.0 -- 1.0 & 0.335\\
      $a$              & \verb|StringZ:aLund|    & 0.0 -- 2.0 & 0.68 \\
      $b$              & \verb|StringZ:bLund|    & 0.2 -- 2.0 & 0.98 \\
      $\left<z_\rho\right>$       & \verb|StringZ:avgZLund| & 0.3 -- 0.7 & (0.55) \\
      \hline
    \end{tabular}
  \end{center}
  \caption{\label{tab:ranges} Parameter ranges used for the \textsc{Pythia} 8 tuning,
    and their corresponding value in the Monash tune. The parenthesis around the Monash value of the $\left<z_\rho\right>$ parameter 
    indicates that this is a derived quantity, not an independent parameter.}
\end{table}

\subsection{Results}

In this section, we discuss briefly the results of the different retunings (for a more detailed discussion please see \cite{Amoroso:2018qga}). In   Table~\ref{tab:T2tune}, we show the results of the tunes with and without the additional $5\%$ flat uncertainty. We can see that the goodness-of-fit is improved a factor of $~7$ bringing it close to unity for the second fit (with $5\%$ uncertainty). Therefore, we can see that the additional $5\%$ uncertainty provide a useful protection against over-fitting.

\begin{table}[!t]
  \begin{center}
    \begin{tabular}{lcc}
      \hline
      Parameter &  without $5\%$ & with $5\%$ \\
      \hline
      \verb|StringPT:Sigma| & $0.3151\substack{+0.0010\\ -0.00010}$ & $0.3227\substack{+0.0028\\ -0.0028}$\\
      \verb|StringZ:aLund|  & $1.028\substack{+0.031\\ -0.031}$ & $0.976\substack{+0.054\\-0.052}$\\
      \verb|StringZ:avgZLund| & $0.5534\substack{+0.0010\\ -0.0010}$ & $0.5496\substack{+0.0026\\ -0.0026}$ \\
      \hline
      $\chi^2/N_\textrm{DoF}$   & 5169/963 & 778/963 \\
      \hline
    \end{tabular}
  \end{center}
    \caption{\label{tab:T2tune} Results of the tunes before and after including a flat $5\%$ uncertainty to the theory prediction.}
\end{table}

\begin{figure}\centering
  \includegraphics[width=0.495\textwidth]{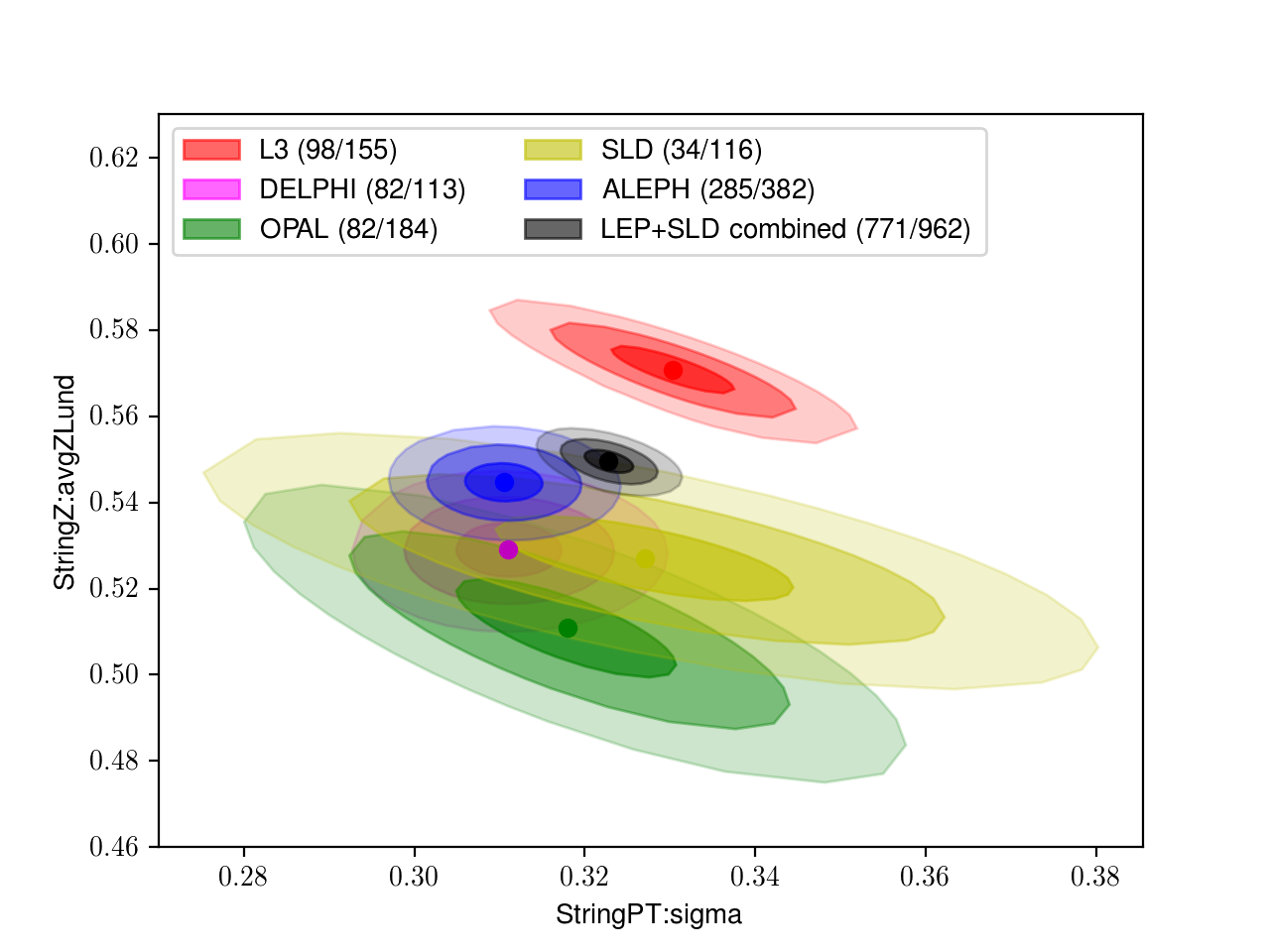}
  \includegraphics[width=0.495\textwidth]{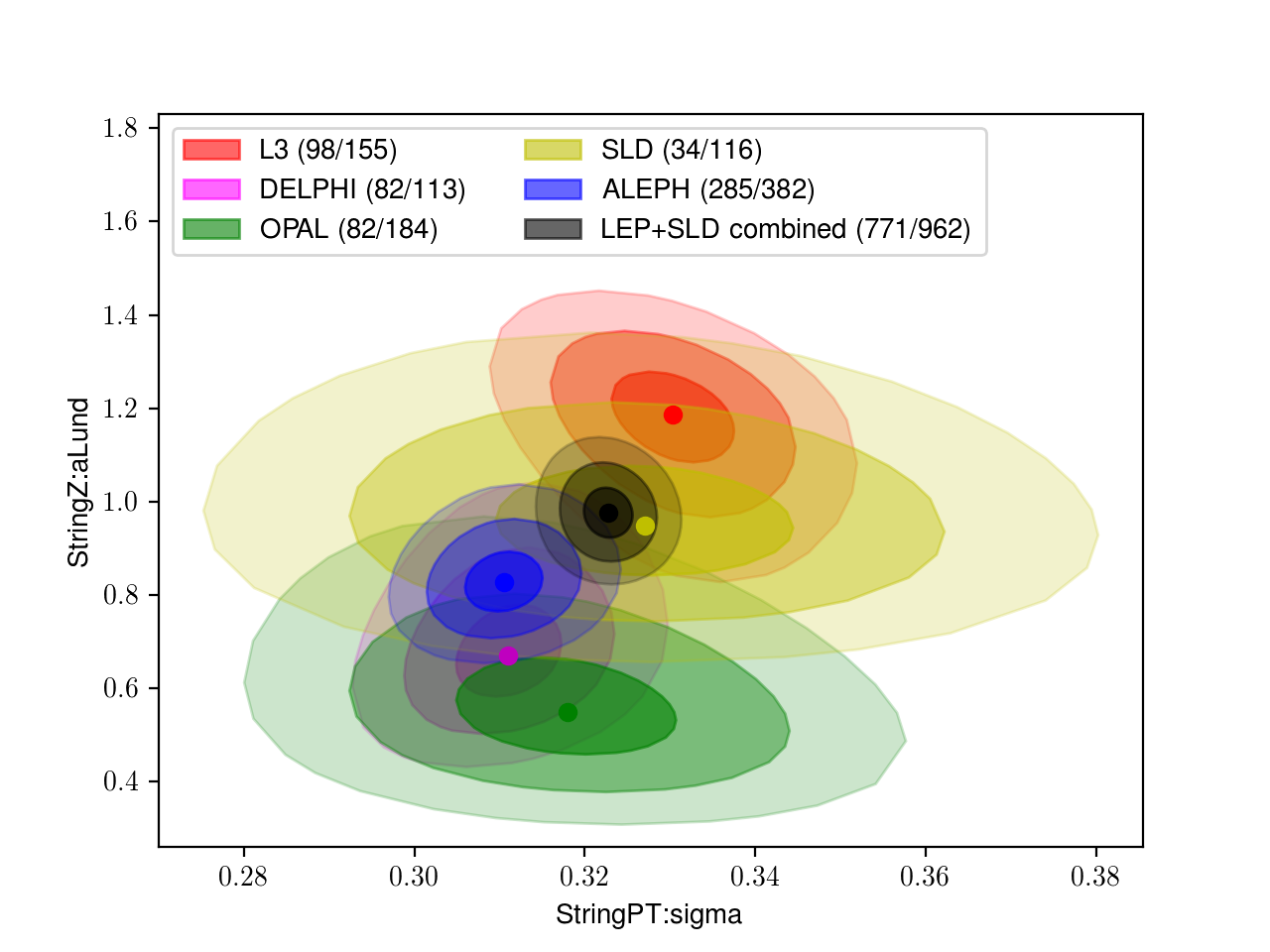}
  \caption{\label{fig:experiments} Results of tunes performed separately to all of the measurements 
  from a given experiment; \textsc{Aleph} (blue), \textsc{Delphi} (magenta), \textsc{L3} (red), \textsc{Opal} (green), \textsc{Sld} (yellow) and 
  COMBINED (gray). The contours corresponding to one, two and three sigma deviations are also shown.}
\end{figure}

We show the possible tensions in the data measured  by the different experiments by making independent tunes including all of the sensitive measurements by each experiment. We performed five independent tunes corresponding to the individual measurements by \textsc{Aleph}, \textsc{Delphi}, \textsc{L3}, \textsc{Opal} and \textsc{Sld} and display these results in Fig. \ref{fig:experiments}. We can see that the tunes to \textsc{Aleph}, \textsc{Delphi}, \textsc{Opal} 
and \textsc{Sld} are in agreement regarding the obtained value of
\verb|StringZ:avgZLund| contrarily to \textsc{L3}. Due to the correlations between the $a$ and the $b$ (or $<z_\rho>$) parameters, we cannot say that these discrepancies in the individual best-fit points is a sign of disagreement between theoretical predictions and data, \emph{i.e.} the predictions at the best-fit point will agree with each other and with data. 

\section{QCD uncertainties}
\label{sec:results}

\subsection{Estimating the uncertainties}
\label{tune:uncertainties}
 
\begin{table}[!t]
 \begin{center}
  \begin{tabular}{lc}
   \hline
   Parameter & Value \\
   \hline
   \verb|StringZ:aLund| & $0.5999\pm0.2$ \\
   \verb|StringZ:avgZLund| & $0.5278^{+0.027}_{-0.023}$ \\
   \verb|StringPT:sigma| & $0.3174^{+0.042}_{-0.037}$ \\
   \hline
  \end{tabular}
 \end{center}
  \caption{\label{tab:individual} Result of the single fit to all the measurements 
  as obtained from independent optimisation to $N(=15)$ measurements. The quoted errors correspond to the $68\%$ CL 
  uncertainty on the fit.}
\end{table}

The QCD uncertainties can be split into two categories: perturbative related to parton showers and non-perturbative related to the hadronisation model parameters. The uncertainties on parton showering within \textsc{Pythia8} were estimated using the automatic method developed in \citep{Mrenna:2016sih}. The uncertainty in this case is determined by variation of the central renormalisation scale by a factor of $2$ in two directions with a full NLO recompensation terms. Furthermore, this framework can allow for variations of the non-singular terms in the DGLAP splitting kernels. We notice that these variations give, in most of the cases, very small uncertainties and, therefore, will be neglected.

On the non-perturbative side, the \textsc{Professor} toolkit allows to estimate uncertainties on the fitted parameters through the \texttt{eigentunes} method which diagonalises the $\chi^2$ covariance matrix around the best-fit point. Then, it uses variations along the principal directions (eigenvectors) in the space of the optimised parameters to construct a set of $2\cdot N_{\textrm{params}}$ variations. However, the resulting \texttt{eigentunes} are found to provide small uncertainties which cannot be interpreted as a conservative\footnote{We have checked that the impact of the \texttt{eigentunes} on the gamma-ray spectra in different final states and for different DM masses including the ones corresponding to the pMSSM best fit points and we have found that the bands obtained from the eigentunes are negligibly small.}. Therefore, we will devise a new method.

The new method consist of making a new tuning where 
we use $N$ different measurements to get $N$ best-fit points. We then
take the $68\%$ CL errors on the parameters to be our estimate of the 
uncertainty (we exclude observables with little or no sensitivity on our parameters).
The results of these fits along with their $68\%$ CL errors are shown
 in Table \ref{tab:individual}. To get a comprehensive estimate of the uncertainty bands from the $68\%$ CL errors on the model parameters, we consider all the possible variations; there are $N_{\mathrm{var}} = 3^3 - 1 = 26$ variations. There are, however, some variations which don't give significant impact on the predicted spectra. We have checked that there are ten variations (including the nominal tune) meaningful variations.

\subsection{Impact on Dark Matter Spectra and Fits}
\begin{figure}[!t]
\centering
\includegraphics[width=0.48\linewidth]{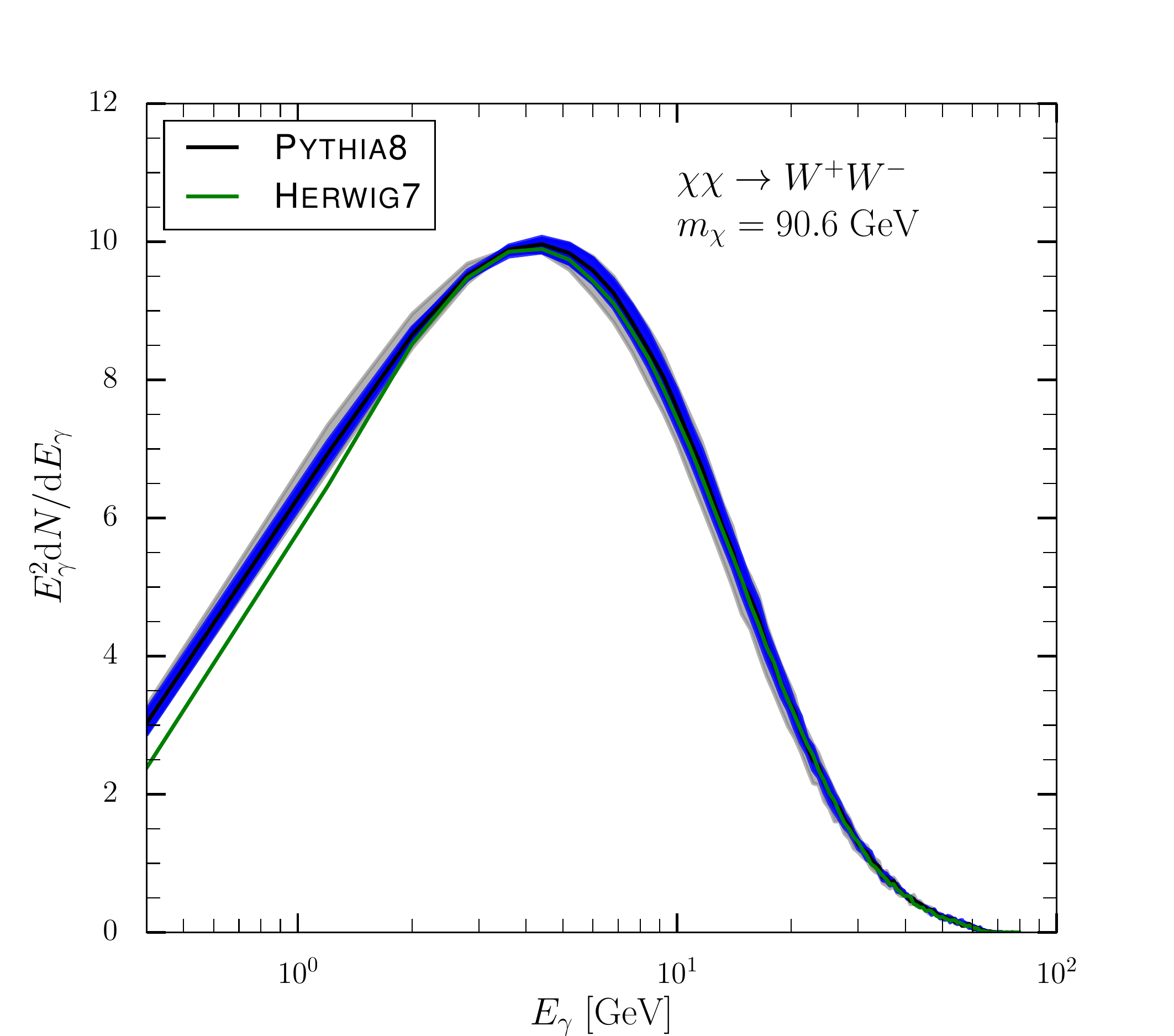}
\hfill
\includegraphics[width=0.48\linewidth]{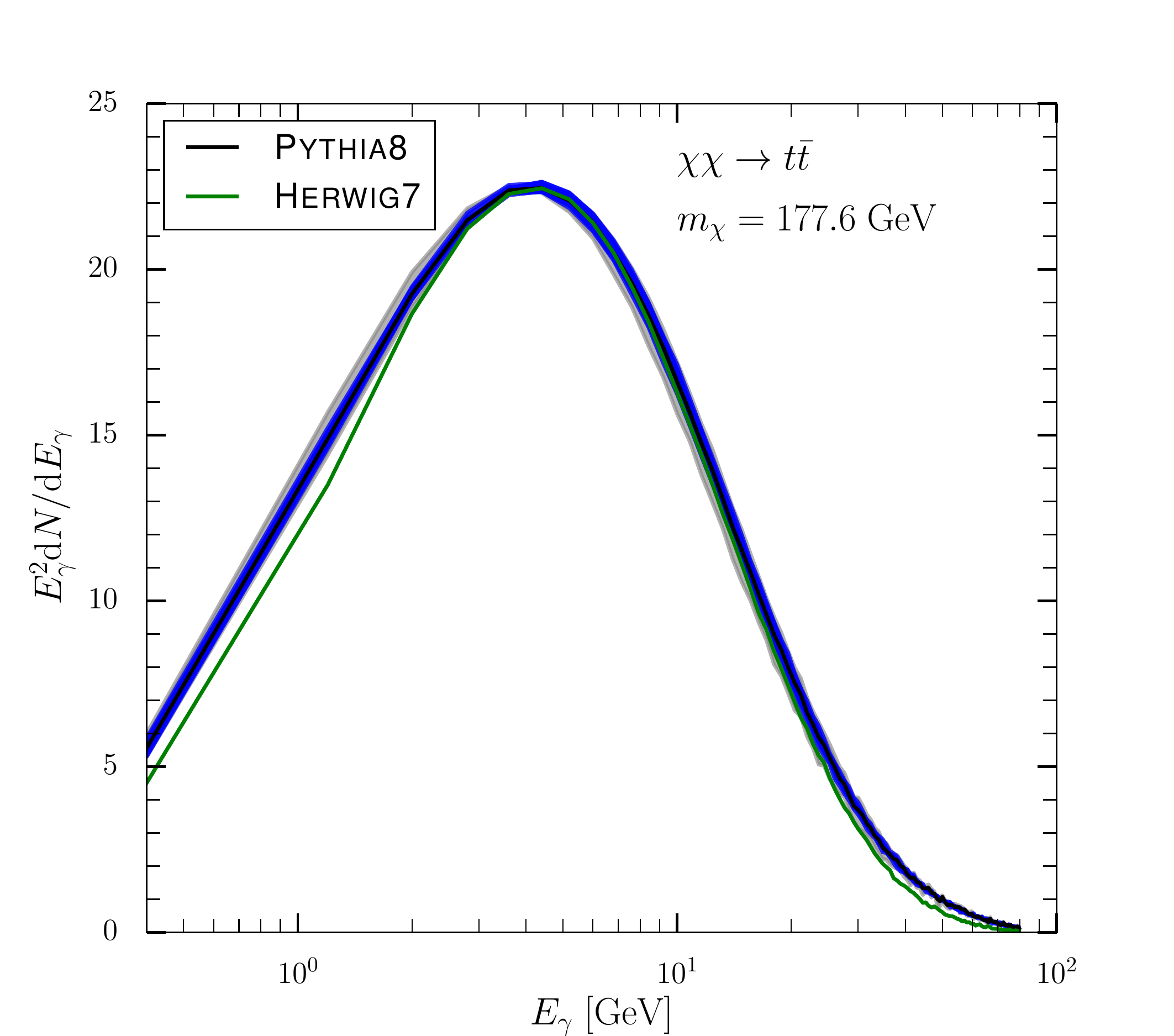}
\caption{Photon energy distribution for dark matter annihilation into $W^+ W^-$ with $m_\chi=90.6$ GeV (\emph{left}) 
and into $t\bar{t}$ with $m_\chi = 177.6$ GeV (\emph{right}). In the two cases, the result corresponding to the new tune is shown in black line. Both the uncertainties from parton showering (gray bands) and from hadronisation (blue bands) are shown. Predictions from \textsc{Herwig7} are shown as a gray solid line.}
\label{fig:DMimpact}
\end{figure}

In this subsection, we show the results of the of QCD fragmentation function, parton shower uncertainties on the photon spectra of two representative DM annihilation channels: $W^+W^-$ and $t\bar{t}$\footnote{For comparison, we show the predictions of \textsc{Herwig7}.}. We do not perform a full analysis to determine the best fit of the GCE, using PASS8 data performed in the pMSSM \cite{Achterberg:2017emt} but only show qualitatively the size of the uncertainties. As the best-fit point will be certainly affected by these uncertainties, we postpone this to a future study. In the analysis of \cite{Achterberg:2017emt}, the best-fit was found for two neutralino masses, i.e $m_\chi=90.6$ GeV and $m_\chi=177.6$ GeV corresponding to the $W^+W^-$ and $t\bar{t}$ DM annihilation channels respectively. These results are shown in Fig. \ref{fig:DMimpact} for $m_\chi=90.6$ GeV in the $W^+W^-$ channel (left panel) and for $m_\chi=177.6$ GeV in the $t\bar{t}$ channel (right panel) with the new tune (black line) and the \textsc{Herwig} prediction (green line). The bands show the \textsc{Pythia} parton-shower (gray bands) and hadronisation (blue bands) uncertainties. We can see that the predictions from \textsc{Pythia} and \textsc{Herwig} agree very well except for $E_\gamma \leqslant 2~$GeV where differences can reach about $21\%$ for $E_\gamma \sim 0.4$ GeV. One can see that the uncertainties can be important for both channels particularly, in the peak region which corresponds to energies where the photon excess is observed in the galactic center region. Indeed, combining them in quadrature assuming the different type of uncertainties are uncorrelated, they can go from few percents where the GCE lies to about 70\% in the high energy bins. Furthermore hadronisation uncertainties are the dominant ones around the peak of the photon spectrum. The parton showering uncertainties can change the peak of the energy spectra and are the main source of uncertainties while moving away toward the edges of the spectra.

\begin{figure}[!t]
\centering
\includegraphics[width=0.32\linewidth]{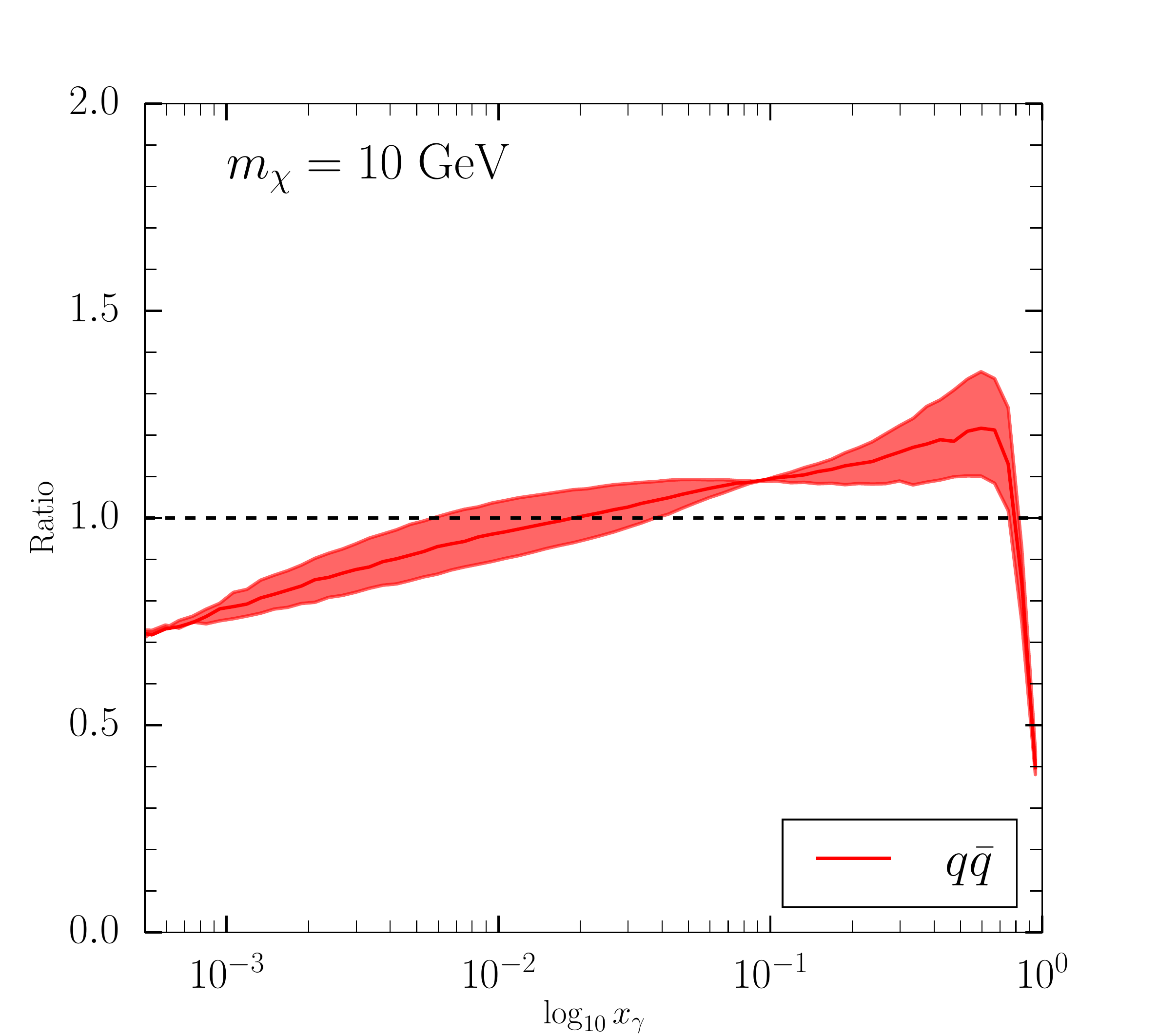}
\hfill
\includegraphics[width=0.32\linewidth]{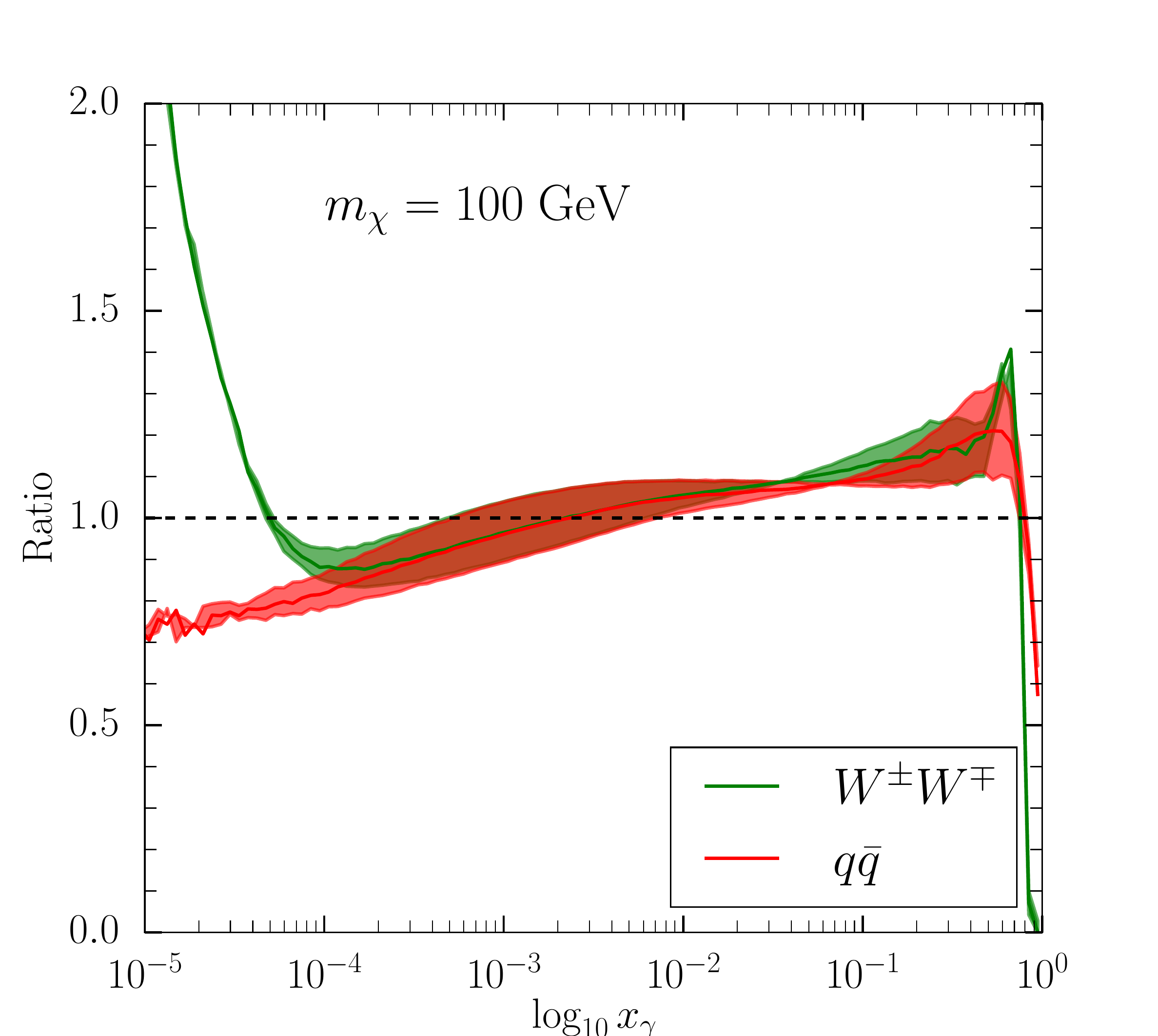}
\hfill
\includegraphics[width=0.32\linewidth]{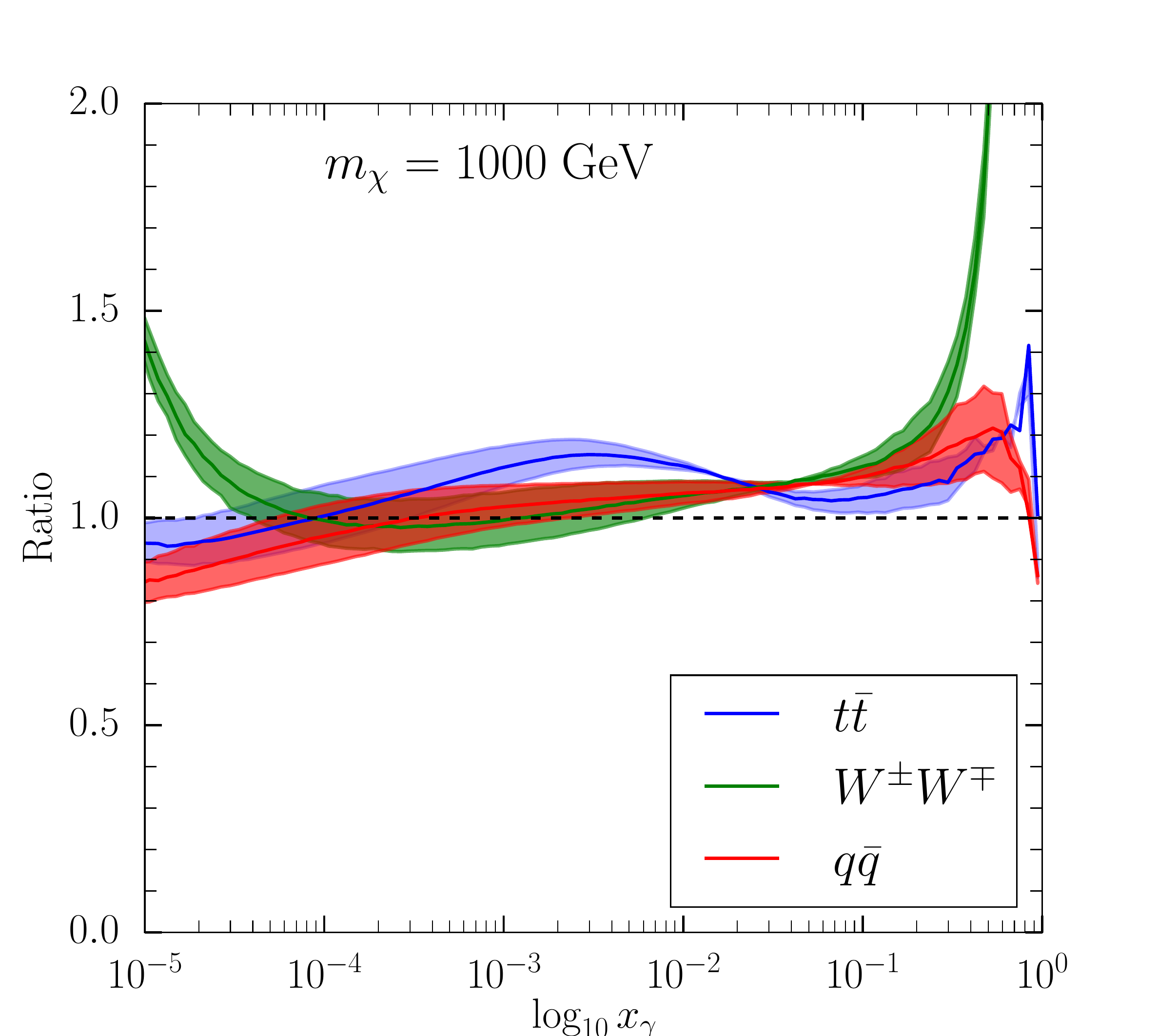}
\caption{Photon spectra obtained using our tune normalized to the results of \cite{Cirelli:2010xx} for $m_\chi = 10$ GeV (\emph{left pane}), $m_\chi = 100$ GeV (\emph{center pane}) and $m_\chi = 1000$ GeV (right pane). The spectra are shown for DM annihilation into $q\bar{q}$ (red), $W^\pm W^\mp$ (green) and $t\bar{t}$ (blue). The dashed bands show the QCD uncertainties on the parameters of the Lund fragmentation function.}
\label{fig:comparison}
\end{figure}

\section{Public Data on Zenodo}
\label{sec:code}


The impact of the QCD uncertainties on the particle spectra from DM annihilation were produced in the form of tables which can be found in Zenodo \cite{Amoroso:2019Zenodo}. We have produced tables for five stables final states; gamma-rays, positrons, electron anti-neutrinos, muon anti-neutrinos and tau anti-neutrinos -- the work on the spectra of anti-protons is in progress \cite{Caron:2021xx} --. The calculations were done for various DM annihilation channels; $\chi\chi \to e^+ e^-, \mu^+ \mu^-, \tau^+ \tau^-, q\bar{q}(q=u,d,s), c\bar{c}, b\bar{b}, t\bar{t}, W^+ W^-, ZZ, gg, ~\mathrm{and}~hh$. We covered DM masses from $5$~GeV to $100~$GeV. For each final state, and annihilation channel, there are twelve tables which are provided in \texttt{zip} format. The notation of the different tables is given below:
\begin{itemize}
    \item The table corresponding to the central prediction for the spectra is denoted by 'AtProduction-Hadronization1-\$TYPE.dat' with \$TYPE=Nuel, Numu, Nuta, Ga, and Positrons refers to the three flavours of anti-neutrinos, gamma-rays and positrons respectively.
     \item There are nine tables corresponding to the different variations of the light quark fragmentation function's parameters. These tables are denoted by 'AtProduction-Hadronization\$h-\$TYPE.dat' with h=2,..,10.
     \item The particle spectra corresponding to the variations of the shower evolution scale ($\mu_R$) are denoted by 'AtProduction-Shower-Var\$s-\$TYPE.dat' with s=1,2  corresponds to $\mu_R/2$ and $2 \mu_R$.   
\end{itemize}
We stress that the uncertainties from parton shower and hadronisation were taken to be uncorrelated (more details on the generation of the spectra can be found in \cite{Amoroso:2018qga}). 

Finally, we have compared our predictions to the results of the PPPC4DMID. We show the comparison between our predictions and the results of the PPPC4DMID in the photon spectra for three DM masses; $m_\chi=10, 100$ and $1000$ GeV. We have chosen three final states, i.e $q\bar{q}, q=u,d,s$, $W^\pm W^\mp$ and $t\bar{t}$. We can see that the differences between our results and the predictions of the Cookbook can be very important, particularly in the edges of the distributions (small $x_\gamma$ and large $x_\gamma$). As these differences cannot be accounted for by QCD uncertainties (shown as dashed bands in Fig. \ref{fig:comparison}), we urge to use the updated predictions from this study. \\

\section{Conclusions}
\label{sec:conclusions}
In this talk, we discussed the study of the QCD uncertainties on particle spectra from DM annihilation which we studied for the \emph{first} time in \cite{Amoroso:2018qga}. We demonstrated that the relative differences between the  predictions of different multi-purposes MC event generators  (\textsc{Herwig} 7.1.3, \textsc{Pythia} 8.235 and \textsc{Sherpa} 2.2.5) cannot be used to define a conservative estimate of QCD uncertainties particularly in the bulk of the spectra. We studied a complementary approach by using the same modeling paradigm (\textsc{Pythia8}) to define parametric variations taking the default \textsc{Monash} tune as our baseline and performed several retunings using data from LEP. Next, we show quantitatively the impact of the QCD uncertainties on the spectra of gamma-rays from DM annihilation in two benchmark points in the pMSSM. Full data tables which can be used to update those in the PPPC4DMID are public now on Zenodo and can be found in \url{http://doi.org/10.5281/zenodo.3764809}.

\end{document}